\documentclass[a4paper,11pt]{article}
\usepackage{amsfonts}
\usepackage{amssymb}
\usepackage{hyperref}
\begin{document}
\title{CMB anisotropy power spectrum of 3-sphere universe for low $l$}
\author{Youngsub Yoon \\\emph{Dunsan-ro 201, Seo-gu} \\\emph{Daejeon 35245, Korea}}
\maketitle

\begin{abstract}
We calculate the CMB anisotropy power spectrum of a closed universe for large angle (i.e., low $l$) due to a scale invariant fluctuation of primordial universe by considering the spherical harmonics for 3-sphere. In particular, contrary to the wide belief, we show that this consideration affects the CMB anisotropy power spectrum; instead of constant $l(l+1)C_l$, our consideration results in the supression for $l=2$, currently explained by the cosmic variance. As a more concrete proof of our analysis, from the low $l$ CMB anisotropy data \emph{alone}, we obtained $\Omega_{\mathrm{tot}}=1.0018^{+0.0031}_{-0.0007}$, which agrees with $\Omega_{\mathrm{tot}}=1.0023^{+ 0.0056}_{- 0.0054}$ from the previous anlaysis of \emph{WMAP}+BAO+$H_0$. Thus, we conclude that our Universe is closed.
\end{abstract}

\section{Introduction}
It is well-known that the CMB anisotropy power spectrum gives us very valuable information about our universe. It is also well-known that the scale-invariance in primordial universe implies a constnat $l(l+1)C_l$ for low $l$. However, observations show that $C_l$ is severely suppressed for $l=2$. This could be due to a big cosmic variance, but it could be due to another effect.

In this article, we will consider a closed universe. Under such a consideration, our universe is necessarily a 3-sphere. In the analysis of CMB anisotropy, only the spherical harmonics of 2-sphere have been considered so far. However, as we will see in this article, considering the spherical harmonics of 3-sphere gives differences to the CMB anisotropy spectrum, even though only for low $l$. In particular, $C_2$ is suppressed, even though not as low as the observed value, implying that the cosmic variance still plays a role, albeit to a less extent.

The organization of this article is as follows. In Section 2, we review the spherical harmonics for 3-sphere. In Section 3, we review how the traditional CMB anisotropy analysis is done by using the spherical harmonics for 2-sphere. The aim is to set a comparison with the case of 3-sphere in Section 4. In Section 4, we derive the CMB anisotropy using the spherical harmonics of 3-sphere. This section is the main part of this paper. In particular, we show that for large $l$, we recover the usual scaling law $l(l+1) C_l=\mathrm{const}$. In Section 5, we use data analysis to obtain the radial distance of the last scattering surface. In particular, we will see that it agrees with the one obtained earlier by another method. 
\section{Spherical harmonics for 3-sphere}
The spherical harmonics on 3-sphere is given by
\begin{equation}
\nabla^2 Y_{qlm}(\chi, \theta,\phi)=-q(q+2)Y_{qlm}(\chi, \theta,\phi)
\end{equation}
where $q$ is a non-negative integer and $l$ runs from 0 to $q$ and $m$ runs from $-l$ to $l$. Of course, we can write
\begin{equation}
Y_{qlm}(\chi,\theta,\phi)=X_{ql}(\chi)Y_{lm}(\theta,\phi)
\end{equation}
for a suitable $X_{ql}$. Given $q$, there is a degeneracy of $(q+1)^2$, as
\begin{equation}
\sum_{l=0}^{q}\sum_{m=-l}^{l}1=\sum_{l=0}^q (2l+1)=(q+1)^2\label{2l+1q+1^2}
\end{equation}

For our purpose, the following relation is important (see \cite{3sphere}, for example)

\begin{equation}
\sum_{lm} Y_{qlm}(\vec u)Y_{qlm}(\vec v)=\frac{q+1}{2\pi^2} U_q(\vec u\cdot \vec v)\label{q+1Uq}
\end{equation}
where $U_q$ is Chebyshev polynomials of the second kind, i.e.,
\begin{equation}
U_q(\cos\theta)=\frac{\sin(q+1)\theta}{\sin\theta}
\end{equation}
and $\vec u$ are $\vec v$ are 4-d unit vectors in unit 3-sphere.
\section{Traditional spherical harmonics analysis in CMB anisotropy}
This section is important to set a comparison with our application of spherical harmonics for 3-sphere.
\begin{equation}
\Delta T(\hat n)=\sum_{lm} a_{lm} Y_l^m(\hat n),\qquad C_l\equiv\langle|a_{lm}|^2\rangle
\end{equation}
\begin{equation}
\langle\Delta T(\hat n)\Delta T(\hat n')\rangle=\sum_{lm} C_l Y_l^m(\hat n)Y_l^{-m}(\hat n')=\sum_l C_l\left(\frac{2l+1}{4\pi}\right) P_l(\hat n\cdot\hat n')\label{Cl2l+1Pl}
\end{equation}
Then, using
\begin{equation}
\int d\Omega_{\hat k} P_l(\hat n \cdot \hat k) P_{l'}(\hat n'\cdot \hat k)=\frac{4\pi}{2l+1} P_l(\hat n\cdot \hat n')\delta_{ll'}\label{PP}
\end{equation}
$C_l$ can be obtained by
\begin{equation}
C_l=\frac{1}{4\pi}\int d^2\hat n d^2\hat n'\, P_l(\hat n \cdot \hat n')\langle\Delta T(\hat n)\Delta T(\hat n')\rangle
\end{equation}
In particular, when Sachs-Wolfe approximation is valid, we can write
\begin{equation}
\frac{\Delta T(\hat n)}{T}=-\frac 15 \mathcal R (\hat n r_L)
\end{equation}
where $\mathcal R$ is the primordial curvature perturbation, and $r_L$ is the radial coordinate of the last scattering surface. When $\mathcal R$ satisfies approximate scale invariance, as widely believed, we have
\begin{equation}
\langle \mathcal R(\lambda \vec x)\mathcal R(\lambda \vec y)\rangle=\langle \mathcal R(\vec x)\mathcal R(\vec y)\rangle
\end{equation}
in which case we have 
\begin{equation}
C_l=\frac{\mathrm{const}}{l(l+1)}
\end{equation}
\section{3d-spherical harmonics in CMB anisotropy} 
Let's re-write (\ref{q+1Uq}) as
\begin{equation}
\sum_{lm} Y_{qlm}(\vec u)Y_{qlm}(\vec v)=\frac{(q+1)^2}{2\pi^2} \left(\frac{U_q(\cos\theta)}{q+1}\right)=\frac{(q+1)^2}{2\pi^2} \left(\frac{\sin(q+1)\theta}{(q+1)\sin\theta}\right)
\end{equation}
where $\cos\theta\equiv \vec u\cdot \vec v$. Then, we can write
\begin{equation}
\langle \Delta T(\hat n) \Delta T(\hat n')\rangle=\sum_q C_q\frac{(q+1)^2}{2\pi^2}\left(\frac{\sin(q+1)\theta}{(q+1)\sin\theta}\right),\qquad C_q\equiv \langle|a_{qlm}|^2\rangle\label{TT3sphere}
\end{equation}
where $\cos\theta$ is the dot product between $\hat n$ and $\hat n'$ in 3-sphere. Here, by an abuse of notation, we denoted the average of $|a_{qlm}|^2$ as $C_q$; this is not the same one as $C_l$.

Now, let's compare this with (\ref{Cl2l+1Pl}). For $\hat n\cdot \hat n'=\cos \theta_{nn'}$, and $\theta_{nn'}$ small, the right-hand side of (\ref{Cl2l+1Pl}) can be expanded as
\begin{equation}
C_l \frac{2l+1}{4\pi} \left(1-\frac{l(l+1)}{4}\theta_{nn'}^2\right)
\end{equation}
In case of (\ref{TT3sphere}) for $\theta$ small, we have
\begin{equation}
C_q \frac{(q+1)^2}{2\pi^2} \left(1-\frac{q(q+2)}{6}\theta^2\right)
\end{equation}
Thus, we see that they indeed have the similar structure. $C_l$ is replaced by $C_q$, the degeneracy $(2l+1)$ is replaced by $(q+1)^2$, the leading term in the parenthesis is both 1, and the coefficients for $\theta_{nn'}^2$ and $\theta^2$ are both proportional to the eigenvalues for the Laplacian. From this reason, we expressed (\ref{TT3sphere}) by pulling out the factor $(q+1)^2$ to the front, instead of the original expression $C_q(q+1)U_q(\cos\theta)/(2\pi^2)$.

Analogous to (\ref{PP}), we have (when $\hat n=\hat n'$)
\begin{equation}
\int_{0}^{\pi}\left(\frac{U_q(\cos\theta)}{q+1}\right)\left(\frac{U_q'(\cos\theta)}{q'+1}\right)\sin^2\theta d\theta\int d\Omega=\frac{2\pi^2}{(q+1)^2}\delta_{qq'}\label{UU}
\end{equation}

In 2-sphere case, we used (\ref{PP}) to obtain $C_l$. However, in 3-sphere case, we shall not use (\ref{UU}), because what we want to obtain is $C_l$ not $C_q$. Moreover, the integration range of (\ref{UU}) is not the subdomain 2-sphere, but the whole 3-sphere, as we can see from the measure $\sin^2\theta d\theta d\Omega$. In other words, we still need to use (\ref{UU}), but only if the integration range is properly considered. As we have
\begin{equation}
\cos\theta=\cos^2\chi+\sin^2\chi \hat n\cdot \hat n'
\end{equation}
the integration range is 
\begin{equation}
\cos 2\chi\leq \cos \theta \leq 1\qquad\longrightarrow \qquad0\leq \theta \leq 2\chi
\end{equation}
  
Thus to obtain $C_l$, we have
\begin{eqnarray}
C_l(2l+1)&=&\sum_{q=l}^{\infty} C_q \frac{(q+1)^2}{2\pi^2}\int_{\cos\theta=\cos2\chi}^{\cos\theta=1} \left(\frac{U_q(\cos\theta)}{q+1}\right)^2\sin^2\theta d\theta \\
&=&\sum_{q=l}^\infty \frac{C_q}{2\pi^2}\int_{0}^{2\chi} d\theta\, \sin^2(q+1)\theta 
\end{eqnarray}
The range for the infinite sum comes from the fact that, for a given $l$, the possible $q$ runs from $l$ to $\infty$. The $(2l+1)$ term in the left-hand side comes from the fact that there is a degeneracy of $(2l+1)$ for a given $l$. In other words, we have $(2l+1)$ factor on the left-hand side and $(q+1)^2$ factor on the right-hand side as expected from (\ref{2l+1q+1^2}).

Now, we need to find $C_q$. Recall that the Lagrangian for $\mathcal R$ in inflation is given by
\begin{equation}
S_2=\frac 12 \int d^3x dt\, 2a^3 \varepsilon \left((\partial_t\mathcal R)^2-\frac{(\partial_i \mathcal R)^2}{a^2}\right),\qquad \varepsilon\equiv -\frac{\dot H}{H^2}
\end{equation}
Considering that $\mathcal R$ is conserved, $\partial_t \mathcal R$ is zero. Thus,
\begin{equation}
\nabla^2_x\langle \mathcal R(x) \mathcal R(y)\rangle=\frac{1}{2a\varepsilon} \delta^3(x-y)
\end{equation}
Therefore, the two-point function of Fourier mode is given by the inverse of the Laplacian. As the eigenvalues of Laplacian is proportional to $q(q+2)$, we conclude $C_q$ is proportional to $1/(q(q+2))$. Thus, for some constant $C$, we have
\begin{eqnarray}
C_l(2l+1)&=&2C \sum_{q=l}^{\infty} \frac{1}{q(q+2)}\left(\chi-\frac{\sin(4(q+1)\chi)}{4(q+1)}\right)\\
&=&C\chi \left(\frac{(2l+1)}{l(l+1)}-\frac {1}{2\chi} \sum_{q=l}^\infty \frac{\sin(4(q+1)\chi)}{q(q+1)(q+2)}\right)\label{Cchi}
\end{eqnarray}
In other words, for some constant $D$, we have
\begin{equation}
l(l+1)C_l=D\left(1-\frac {l(l+1)}{2\chi(2l+1)} \sum_{q=l}^\infty \frac{\sin(4(q+1)\chi)}{q(q+1)(q+2)}\right)\label{D}
\end{equation}
Thus, we see that the second term gives the deviation from the usual Sachs-Wolfe plateau. However, this term rapidly converges to zero for higher $l$, not only because there are fewer terms to add (even though there are infinite terms to do so), but also because the sine function is oscillating.

\section{Data analysis}
Let's set the notation. We have
\begin{equation}
\chi=\sqrt{\Omega_{\mathrm{tot}}-1}\int_{1/(1+z_L)}^1\frac{da}{a^2\sqrt{\Omega_{\Lambda}-(\Omega_{\mathrm{tot}}-1)a^{-2}+\Omega_M a^{-3}}},\qquad r_L=\sin \chi
\end{equation}
where we ignored the contribution from the radiation. We use $\Omega_M=\Omega_{\mathrm{tot}}-\Omega_\Lambda$.

Let's see what the previous analysis yields for $\chi$. From \emph{WMAP}+BAO+$H_0$, they obtained \cite{Jarosik}
\begin{equation}
\Omega_{\mathrm{tot}}=1.0023^{+ 0.0056}_{- 0.0054},\quad \Omega_\Lambda=0.728^{+0.015}_{-0.016},\quad z_L=1090.89^{+ 0.68}_{- 0.69}\label{Otot}
\end{equation}
which yields, according to our calculation,
\begin{equation}
\chi=0.16^{+0.14}_{-0.16}\label{chiwmap}
\end{equation}

Now, it's our turn to calculate $\chi$ by using low $l$ CMB anisotropy data in \cite{CMB Data}. As the present author does not know well about statistics and data processing, we tried to find the best fit $\chi$ by trial and error. First, we define $D_l\equiv C_l\cdot l(l+1)$. Then, notice
\begin{equation}
\lim_{l\to \infty} D_l=D
\end{equation}
By averaging $D_l$ from $l=2$ to 29, we get $D=851$. Then, we tried to minimize 
\begin{equation}
\frac{(a_{\mathrm{obs}}-a_{\mathrm{th}})^2}{\sigma_a^2}+\frac{(b_{\mathrm{obs}}-b_{\mathrm{th}})^2}{\sigma_b^2}+\frac{(c_{\mathrm{obs}}-c_{\mathrm{th}})^2}{\sigma_c^2}\label{abc}
\end{equation} 
where ``obs'' denotes observed value, and ``th'' denotes theoretical value, and
\begin{equation}
a=C_2\cdot 2\cdot 3,\qquad b=C_3 \cdot 3\cdot 4,\qquad  c=\frac{(C_4\cdot 4\cdot 5)+(C_5\cdot 5\cdot 6)}{2}
\end{equation}
$\sigma$s are also in the Table 1.

We found that $\chi=0.14$ minimizes (\ref{abc}). To determine the error of $\chi$, notice that observation \cite{CMB Data} shows that $l=2$ is suppressed while $l=5$ is augmented. This is true for $\chi=0.11$ to 0.23. Therefore, we conclude $\chi=0.14^{+0.09}_{-0.03}$ which agrees with (\ref{chiwmap}). Put it differently, we obtain $\Omega_{\mathrm{tot}}=1.0018^{+0.0031}_{-0.0007}$, which agrees with (\ref{Otot}).
\begin{table}
	\caption{$D_l(\chi)$}
	\centering
	\begin{tabular}{|l|l|l|l|l|l|}
		\hline
		$l$ & $D_{l\mathrm{obs}}$ & $D_{l\mathrm{th}}(0.11)$& $D_{l\mathrm{th}}(0.14)$  & $D_{l\mathrm{th}}(0.16)$ & $\sigma_{D_l}$ \\ \hline
		2 & 150 & 562& 664 & 721 & 708 \\
		3 & 902 & 705 & 800 & 844 & 565 \\
		4.5 & 1099 & 833 & 887 &  899 & 312 \\
		\hline
	\end{tabular}
\end{table}
\section{Discussions and Conclusions}
In this article, we successfully examined the CMB anisotropy power spectrum by a novel approach. As mentioned, we explained the suppresion of $C_l$ for $l=2$, and found an agreement for the value of $\chi$ with the one from the previous analysis. Here, we want to emphasize again that we obtained $\chi$ from by using \emph{only} low $l$ data, while the previous analysis obtained this value by using the whole range of data. This agreement is highly non-trivial. Had the CMB anisotropy observation data showed a ``perfect'' Sachs-Wolfe plateau (i.e., constant $l(l+1) C_l$) for low $l$ (i.e., in our analysis, $l=2$ to 5), as the consideration of spherical harmonics for 2-sphere predicts, we would have obtained $\chi=\infty$ by considering the 3-sphere spherical harmonics. There would have been no agreement.

In conclusion, the spherical harmonics of 3-sphere explain the CMB anisotropy data well, and allow us to calculate $\Omega_{\mathrm{tot}}$, which clearly shows that our Universe is closed;  the spherical harmonics of the 3 pseudo-sphere is qualitatively different, with a completely different prediction for the CMB anisotropy. Future work should perform our numerical analysis in this article again with more rigor. 
\begin{center}
	{\Large\bf Acknowledgments}
\end{center}
I thank Donghui Jeong for help.


\begin{thebibliography}{9}

\bibitem{3sphere}
``Wavelets on the Three-Dimensional Sphere $S^3$,''
Svend Ebert, Diploma Thesis

\url{http://www.mathe.tu-freiberg.de/inst/amm1/Mitarbeiter/Ebert/Wavelets.pdf}

\bibitem{Jarosik} 
N.~Jarosik {\it et al.},
``Seven-Year Wilkinson Microwave Anisotropy Probe (WMAP) Observations: Sky Maps, Systematic Errors, and Basic Results,''
Astrophys.\ J.\ Suppl.\  {\bf 192}, 14 (2011)
doi:10.1088/0067-0049/192/2/14
[arXiv:1001.4744 [astro-ph.CO]].

\bibitem{CMB Data}
\url{https://lambda.gsfc.nasa.gov/data/map/dr5/dcp/spectra/wmap_tt_spectrum_9yr_v5.txt}

\url{https://lambda.gsfc.nasa.gov/data/map/dr5/dcp/spectra/wmap_binned_tt_spectrum_9yr_v5.txt}
\end{thebibliography}
\end{document}